\documentclass[twocolumn,showpacs,preprintnumbers,amsmath,amssymb,prl,superscriptaddress]{revtex4-1}
\usepackage{bbm}
\usepackage{mathrsfs}
\usepackage{graphicx}
\usepackage{dcolumn}
\usepackage{bm}
\usepackage{amsmath}
\usepackage{amsfonts}
\usepackage{color}
\usepackage[colorlinks=true,linkcolor=blue]{hyperref}
\usepackage{floatrow}

\begin{document}

\title{Antiferromagnetic Topological Nodal Line Semimetals}
\author{Jing Wang}
\affiliation{State Key Laboratory of Surface Physics and Department of Physics, Fudan University, Shanghai 200433, China}
\affiliation{Collaborative Innovation Center of Advanced Microstructures, Nanjing 210093, China}

\begin{abstract}
We study three-dimensional nodal line semimetals (NLSMs) with magnetic ordering and strong spin-orbit interaction. Two distinct classes of magnetic NLSMs are proposed. The first class is band-inversion NLSM where the accidental line node is induced by band inversion and locally protected by glide mirror plane and the combined time-reversal and inversion symmetries. This can be viewed as a trivial stacking of the two-dimensional antiferromagnetic Dirac semimetals. The second class is essential NLSM where the nodal features are filling-enforced by specific magnetic symmetry group. We further provide two concrete tight-binding models for magnetic NLSMs which belong to these two different classes, respectively. We conclude with a brief discussion on the possible material venues and the experimental implications for such phases.
\end{abstract}

\date{\today}


\maketitle

The discovery of the time-reversal ($\Theta$) invariant topological insulators (TIs)~\cite{hasan2010,qi2011,chiu2016,bansil2016} has inspired intense research interest in topological semimetals. They are characterized by the point nodes or nodal lines (NLs) where the conduction and valence bands cross in the Brillouin zone (BZ). The nodal point semimetals have linear energy dispersions along all momentum directions around the point nodes, which can be classified in terms of the node degeneracies, including Dirac semimetal (DSM), Weyl semimetal (WSM), Double DSM and Spin-$1$ WSM~\cite{horava2005,young2012,wangzj2012,wangzj2013,young2014,yang2014,liu2014a,liu2014b,borisenko2014,xiong2015,murakami2007,wan2011,xu2011,burkov2011,liu2014,huang2015,weng2015,xu2015,lv2015,bernevig2015,sun2015,ruan2016,deng2016,wieder2016,bradlyn2016}. The nodal line semimetals (NLSMs) have line band crossing with no dispersion along NL direction and linear dispersion in the perpendicular directions~\cite{burkov2011b,chiu2014,weng2015b,kim2015,yu2015,fang2015,xie2015,meng2015,li2016,huang2016,bian2016,mao2016,schoop2016,kaminski2016,wan2017,hirayama2017,xu2017,murakami2017,yang2017,mullen2015,kee2012,chen2015,liang2016,soluyanov2016}, which are protected by the combination of exact crystalline symmetries and topology~\cite{kim2015,yu2015,fang2015}. Drumheadlike surface flat bands are predicted to exit in NLSM~\cite{weng2015b,kim2015,yu2015}, which may lead to high-temperature superconductivity~\cite{volovik2011a}. With certain symmetry breaking, NLSM will evolve into various exotic topological states such as TI and nodal point semimetals. The earlier studies on NLSM materials are focused in $\Theta$-invariant systems without~\cite{weng2015b,kim2015,yu2015,fang2015,xie2015,meng2015,li2016,huang2016,bian2016,mao2016,schoop2016,kaminski2016,wan2017,hirayama2017,xu2017,murakami2017} and with~\cite{kee2012,chen2015,liang2016,soluyanov2016} spin-orbit coupling (SOC).

The $\Theta$-symmetry breaking in general will destroy the robustness of NLs. This motivates us to study possible NLSMs in magnetic systems, which may provide a platform for the interplay between magnetism and exotic topological states. The goal in this paper is to explain how a magnetic NLSM with broken $\Theta$-symmetry and strong SOC can nevertheless exist in three dimensions (3D). Similar to DSM~\cite{parameswaran2013,watanabe2015,po2016,watanabe2016,wieder2016b,chen2016,young2015,wang2017a,tang2016,young2016,wang2017b}, there are two distinct classes of NLSM in magnetic systems. The first class is \emph{band-inversion} NLSM where the line nodes are accidental and intimately related to 2D AFM Dirac points (DPs). Conversely, the second class is \emph{essential} NLSM where the NLs are filling-enforced by specific space group (SG) symmetries. Two tight-binding models are provided for magnetic NLSMs which belong to these two different classes, respectively. We conclude with a brief discussion on the surface states, the possible material venues and the experimental implications for such phases.

\begin{figure}[b]
\begin{center}
\includegraphics[width=3.3in]{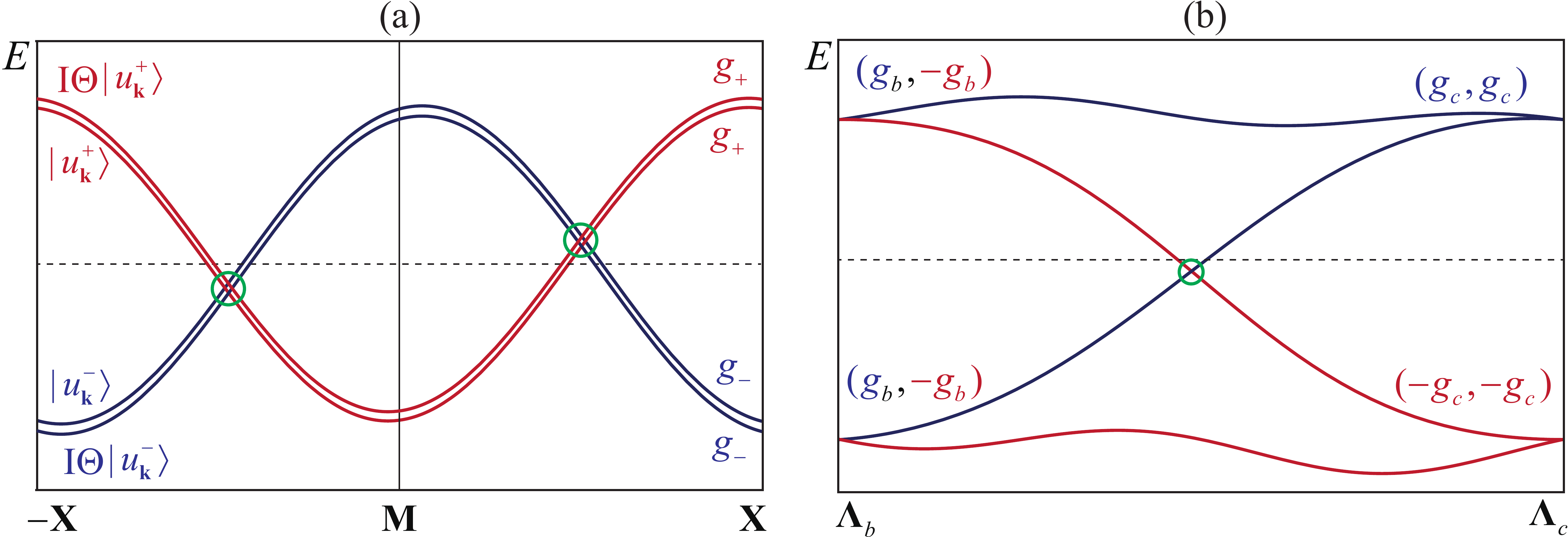}
\end{center}
\caption{(color online) AFM NLSMs. (a) Band-inversion NLSM. Without $\mathcal{I}$ and $\Theta$, but with combined $\mathcal{I}\Theta$, the system is doubly degenerate, which is artificially split for clarity. Along \emph{any} direction in the mirror invariant plane of the BZ, the Bloch states $(|u^+_{\mathbf{k}}\rangle,\mathcal{I}\Theta|u^+_{\mathbf{k}}\rangle)$ and $(|u^-_{\mathbf{k}}\rangle,\mathcal{I}\Theta|u^-_{\mathbf{k}}\rangle)$ have opposite glide eigenvalues, if they cross, then the crossing points (indicated by green circles) are robust. (b) Essential NLSM. The schematic of energy bands along an arbitrary path $\mathcal{C}$ connecting $\boldsymbol{\Lambda}_{b,c}$ on the mirror invariant plane. The Kramers pairs exchange the glide eigenvalues along $\mathcal{C}$, leads to an essential crossing point at $\mathbf{k}_{\mathcal{C}}$.}
\label{fig1}
\end{figure}

\emph{Band-inversion NLSM.}
In a 3D $\Theta$-invariant system without SOC, NLSM emerges through band inversion transitions and is topologically protected by combined inversion ($\mathcal{I}$) and $\Theta$ with $(\mathcal{I}\Theta)^2=+1$~\cite{kim2015,yu2015,fang2015}. Such $\mathcal{I}\Theta$ symmetry also guarantees the 2D DPs in graphene. In the presence of SOC, $(\mathcal{I}\Theta)^2=-1$ and the NL is no longer protected by $\mathcal{I}\Theta$ without additional crystalline symmetries. This can be easily seen by considering the generic 4-band effective Hamiltonian respecting $\mathcal{I}\Theta$ symmetry as $\mathcal{H}(\mathbf{k})=\sum_{a=1}^5d_a(\mathbf{k})\Gamma^a$, where $\Gamma^a$ are $4\times4$ matrices satisfying $\{\Gamma^a,\Gamma^b\}=2\delta_{ab}$, and the specific form of $\Gamma^a$ depends on the representation of $\mathcal{I}\Theta$ with $[\mathcal{I}\Theta,\Gamma^a]=0$. $d_a(\mathbf{k})$ are real and even functions of $\mathbf{k}=(k_x,k_y,k_z)$. The energy spectrum is $E_{\pm}(\mathbf{k})=\pm\sqrt{\sum_{a=1}^5d_a^2(\mathbf{k})}$.
The NLSM can be generated when all $d_a(\mathbf{k})=0$ for a \emph{line} of $\mathbf{k}$ in the BZ, which is only possible when certain crystalline symmetry sets further constraints on $d_a(\mathbf{k})$. For each plane of BZ crossing such NL, the system is a 2D DSM. Therefore, the symmetry classification and protection of NLSM in 3D is similar to that of DSM in 2D~\cite{young2015,wang2017a}. This connection suggests a route towards realizing the NLSM in 3D antiferromagnetic (AFM) systems: starting with a 2D AFM DSM, stack them along $z$-direction while preserving the crystalline symmetries. The symmetry-protected DPs in 2D will evolve into protected NLs in 3D.

\begin{figure}[b]
\begin{center}
\includegraphics[width=3.3in]{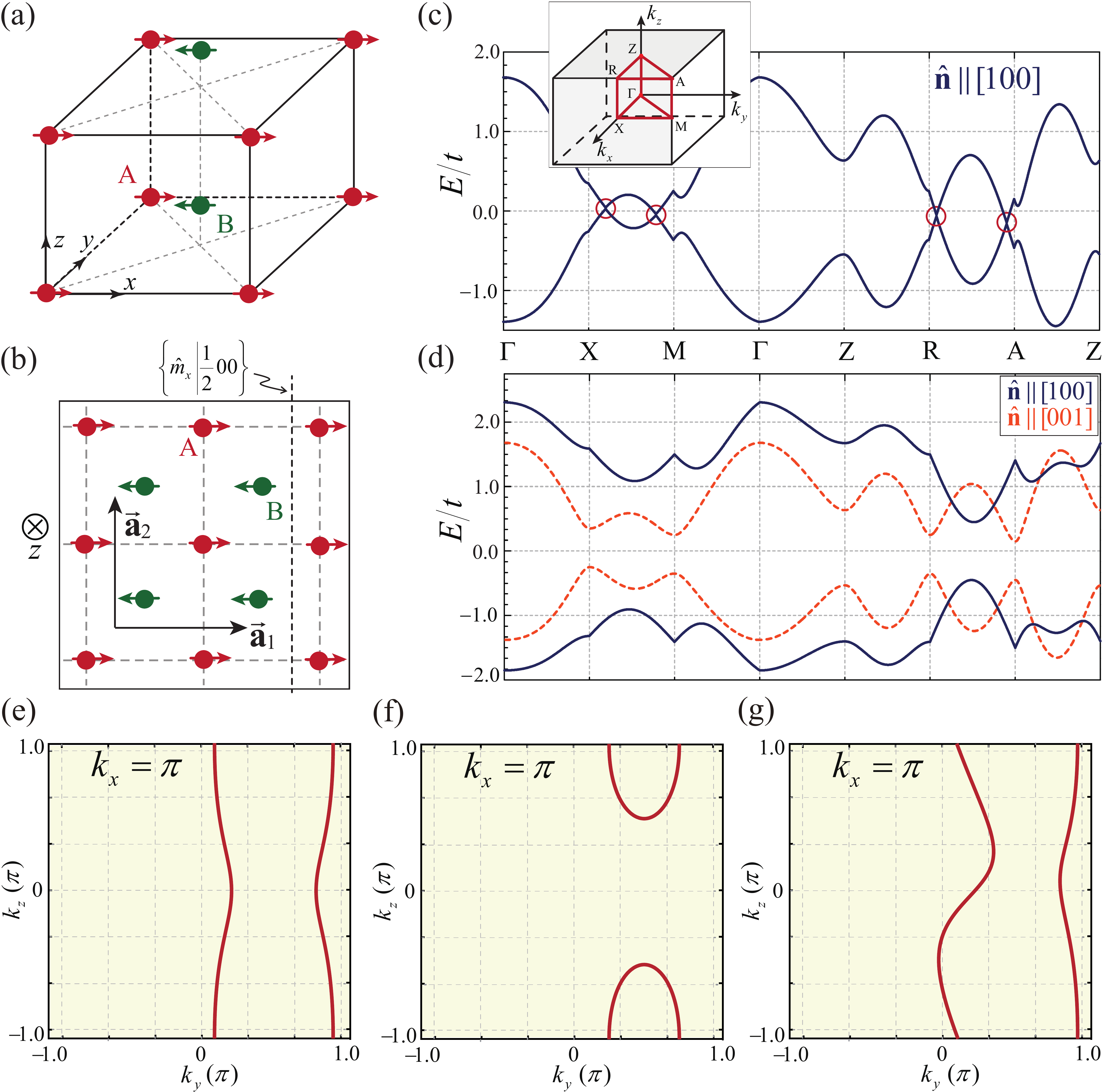}
\end{center}
\caption{(color online). (a) The tetragonal primitive lattice structure of SG~59. The magnetic moments are along $\pm\hat{x}$ direction. (b) Top view of the crinkled 2D square lattice respecting the glide plane $\{\hat{m}_x|\frac{1}{2}00\}$. (c) Energy band along high-symmetry lines of the BZ (inset), described by $\mathcal{H}_a$ in Eq.~(\ref{model1}), with $t_{xy}=1.0$, $t_z=0.5$, $t_{xy}'=t_z'=0.05$, $\lambda=0.8$, $\lambda_z=0.3$, $\Delta=0.3$, and $\hat{\mathbf{n}}\parallel[100]$. The 2D DPs are indicated by red circles. The Dirac NLs on $k_x=\pi$ are shown in (e) $\Delta=0.3$ and (f) $\Delta=0.8$. (d) $\hat{\mathbf{n}}\parallel[001]$ with $\Delta=0.3$ (dashed line), the system is fully gapped, which breaks $\{\hat{m}_x|\frac{1}{2}00\}$ but preserves $\{\hat{C}_{2x}|\frac{1}{2}00\}$, $\{\hat{C}_{2y}|0\frac{1}{2}0\}$. While $\hat{\mathbf{n}}\parallel[100]$ with $\Delta=1.3$ (solid line), the NLs disappear even $\{\hat{m}_x|\frac{1}{2}00\}$ is present. (f) Line nodes for $\mathcal{H}_a+\mathcal{H}_1^a$, with $\Delta=0.3$, $\lambda_1=0.4$, and $\lambda_2=0.2$.}
\label{fig2}
\end{figure}

Nonsymmorphic symmetries can stick bands together and lead to extra degeneracies~\cite{bradley1972}, which is a simple consequence of the noncommutativity of symmetry operators. Here we show for example $\mathcal{I}\Theta$ and glide mirror plane $\mathcal{G}_x\equiv\{\hat{m}_x|\mathbf{t}\}$ could protect the NLs at BZ boundary. $\mathbf{t}=(t_x,t_y,t_z)$ is a fractional primitive lattice vector. The mirror invariant plane includes $k_x=0,\pi$. $\mathcal{G}_x^2=-e^{-i2\mathbf{k}\cdot\mathbf{t}_{\parallel}}$, where $\mathbf{t}_{\parallel}$ is projection of $\mathbf{t}$ in the mirror plane and the minus sign is from equivalent $2\pi$ rotation of spins. The Bloch states at $k_x=0,\pi$ are eigenstates of $\mathcal{G}_x$ with eigenvalues $g_\pm=\pm ie^{-i\mathbf{k}\cdot\mathbf{t}_{\parallel}}$. Now suppose a Bloch state $|u^+_\mathbf{k}\rangle$ at $k_x=0,\pi$ has eigenvalue $g_+$, then its degenerate partner $\mathcal{I}\Theta|u^+_{\mathbf{k}}\rangle$ is also eigenstate of $\mathcal{G}_x$ with eigenvalue $e^{-2ik_xt_x}g_-$.  $\mathcal{G}_x\mathcal{I}\Theta=e^{-i2\mathbf{k}\cdot\mathbf{t}}\mathcal{I}\Theta\mathcal{G}_x$. Therefore at $k_x=0$, $|u^+_{\mathbf{k}}\rangle$ and $\mathcal{I}\Theta|u^+_{\mathbf{k}}\rangle$ have opposite $\mathcal{G}_x$ eigenvalues, any band crossing is generally unstable; while at $k_x=\pi$, the degenerate bands have same $\mathcal{G}_x$ eigenvalues only when $t_x=1/2$. In this case, two sets of degenerate bands with opposite $\mathcal{G}_x$ eigenvalues may cross each other along a NL as shown in Fig.~\ref{fig1}(a), which must be robust. However, such crossing is optional feature of the crystalline symmetry, which is further seen by studying the effective model at $\mathbf{k}=(\pi00)$. Explicitly, we set $\mathcal{I}\Theta=i\sigma_2\mathcal{K}$, $\Gamma^{1,2,3,4,5}=(\tau_1,\tau_2\sigma_3,\tau_2\sigma_1,\tau_2\sigma_2,\tau_3)$, $\tau_i$ and $\sigma_i$
are Pauli matrices acting on the orbital and spin, respectively. $\mathcal{G}_x=\tau_3$ and constrains $d_{1,2,3,4}(k_x,\mathbf{k}_{\parallel})=-d_{1,2,3,4}(-k_x,\mathbf{k}_{\parallel})$, $d_{5}(k_x,\mathbf{k}_{\parallel})=d_{5}(-k_x,\mathbf{k}_{\parallel})$. Therefore, at $k_x=\pi$, only $d_5$ term survives which is an even function of $\mathbf{k}$. To the lowest order, $d_5(\mathbf{k})=m-b\mathbf{k}^2_{\parallel}$. $d_5(\mathbf{k})=0$ is satisfied only when $mb>0$, which is simply the band inversion condition.

We now construct a simple tight-binding model which exhibits the above behavior. The lattice has a tetragonal primitive structure of SG~59 (layer group $Pmmn$) as shown in Fig.~\ref{fig2}(a). The lattice vectors are $\vec{\mathbf{a}}_1=(100)$, $\vec{\mathbf{a}}_2=(010)$, $\vec{\mathbf{a}}_3=(001)$. The system forms a layered structure with two sublattices in one unit cell labeled by $A$ and $B$, and the AFM ordering is along $\hat{\mathbf{n}}$ direction. Each 2D layer has a crinkled square lattice~\cite{wang2017a} with $A$ and $B$ shifting along $z$-axis with the displacement $r_{AB}=(\frac{1}{2}\frac{1}{2}c)$, where $0<c<1/2$. Each lattice site contains an $s$ orbital, which leads to a 4-band model. The system breaks $\mathcal{I}$ and $\Theta$ but respects $\mathcal{I}\Theta$. The Hamiltonian is
\begin{eqnarray}\label{model1}
\mathcal{H}_a &=& \left[t_{xy}\tau_1+t_z\left(\tau_1\cos k_z+\tau_2\sin k_z\right)\right]\cos\frac{k_x}{2}\cos\frac{k_y}{2}
\nonumber
\\
&&+t'_{xy}(\cos k_x+\cos k_y)+t_z'\cos k_z+\Delta\tau_3\boldsymbol{\sigma}\cdot\hat{\mathbf{n}}
\nonumber
\\
&&+\left(\lambda-\lambda_z\cos k_z\right)\tau_3\left(\sigma_2\sin k_x-\sigma_1\sin k_y\right).
\end{eqnarray}
Here $t_{xy}$ and $t'_{xy}$ describe the intra-layer nearest- and next-nearest-neighbor hopping, respectively. $t_z,t_z'$ describe the inter-layer hoppings. $\lambda,\lambda_z$ is SOC and $\lambda>\lambda_z$. $\Delta$ denotes the AFM exchange coupling.

The symmetry of the system depends on the magnetic order direction. If $\hat{\mathbf{n}}\parallel[100]$, it preserves $\{\hat{m}_x|\frac{1}{2}00\}$ and $\{\hat{m}_z|\frac{1}{2}\frac{1}{2}0\}$. As shown in Fig.~\ref{fig2}(c), the band is inverted at the $k_x=\pi$ plane, forming 2D DPs on the $X$-$M$ and $R$-$A$ lines. These DPs signal the presence of NLs on $k_x=\pi$, which is illustrated as red curves on the BZ boundary plane as shown in Fig.~\ref{fig2}(e). The NLs are located at $(\pi,k_y,k_z)$ satisfying $\sin k_y=\Delta/(\lambda-\lambda_z\cos k_z)$, which is mirror symmetric with respect to $k_y=\pi/2$ and $k_z=\pi$. The position and topology of NLs is tunable by $\Delta$, evolving from two unlinked open lines when $\left|\Delta\right|<\lambda-\lambda_z$ to a single line when $\lambda-\lambda_z<|\Delta|<\lambda+\lambda_z$. The Berry phase along a path enclosing each of these NLs is found to be $\pi$, which ensures the topological stability of the NLs. Take Fig.~\ref{fig2}(e) for example, the NL is dispersionless when $\lambda_z=0$. For each $k_x$-$k_y$ plane with fixed $k_{z}=k_{z0}$, the system is a AFM DSM consisting two DPs at $\mathbf{k}_1=(\pi,k_{y0},k_{z0})$ and $\mathbf{k}_2=(\pi,\pi-k_{y0},k_{z0})$, where $\sin k_{y0}=\Delta/(\lambda-\lambda_z\cos k_{z0})$. The effective model near $\mathbf{k}=\mathbf{k}_1$ is $\mathcal{H}_a(\mathbf{k}_1+\mathbf{q})= \left(v_1\tau_1+v_2\tau_2+v_3\tau_3\otimes\sigma_2\right)q_x
+\left(v_4q_y+v_5q_z\right)\tau_3\otimes\sigma_1$, where $v_i$ is obtained from Eq.~(\ref{model1}).
This is a Dirac Hamiltonian, which features a Dirac NL at $q_x=0$ and $v_4q_y+v_5q_z=0$. The NLs are symmetry-protected. $\mathcal{I}\Theta=i\tau_1\sigma_2\mathcal{K}$ allows the mass terms $\tau_2$ and $\tau_3\sigma_3$, which is forbidden by $\{\hat{m}_x|\frac{1}{2}00\}=i\tau_3\sigma_1$. We can further break $\{\hat{m}_z|\frac{1}{2}\frac{1}{2}0\}$ but keep $\{\hat{m}_x|\frac{1}{2}00\}$ by shifting $B$ sites along $y$ axis~\cite{wang2017a}, which allows terms
\begin{eqnarray}\label{model1_add}
\mathcal{H}^a_1 &=& \left[t_1\tau_2+t_2\left(\tau_2\cos k_z-\tau_1\sin k_z\right)\right]\cos\frac{k_x}{2}\sin\frac{k_y}{2}
\nonumber
\\
&&+\left[\lambda_1+\lambda_2\left(\cos k_x+\cos k_y\right)\right]\sin k_z\tau_3\sigma_1
\nonumber
\\
&&+\left(\lambda_3+\lambda_2\cos k_z\right)\sin k_x\tau_3\sigma_3.
\end{eqnarray}
The NLs located at $k_x=\pi$ remain protected as shown in Fig.~\ref{fig2}(g), consistent with the above analysis that $\mathcal{I}\Theta$ and $\{\hat{m}_{\boldsymbol{\ell}}|\mathbf{t}\}$ with $\boldsymbol{\ell}\cdot\mathbf{t}=0$ cannot protect the NLs. However, it is noted the NLs here are \emph{nonessential}, which is only locally permitted by crystalline symmetries.
As shown in Fig.~\ref{fig2}(d), when AFM interaction is much stronger than hopping and SOC terms, even though $\{\hat{m}_x|\frac{1}{2}00\}$ and $\mathcal{I}\Theta$ are present, the Dirac line nodes disappear and the system is full gapped.

\emph{Essential NLSM.} Then we turn to essential NLSM, which is distinct from band-inversion NLSM in that the NLs are guaranteed to exist at Fermi level by certain electron filling. The relationship between filling and the essential nodal points has been studied in $\Theta$-invariant~\cite{parameswaran2013,watanabe2015,po2016,watanabe2016,wieder2016b,chen2016,young2015} and $\Theta$-broken~\cite{young2016,wang2017b} systems. The essential NLs are also studied in $\Theta$-invariant spinful~\cite{chen2015,soluyanov2016,chen2016} and spinless~\cite{murakami2017} systems. Here we consider the symmetry mechanism for essential NLs with $\Theta$-breaking. Similar to the 3D essential magnetic DSM~\cite{wang2017a}, the key point is rooted in a new antiunitary operator $\bar{\Theta}=\Theta T_{\mathbf{d}}$ or $\bar{\Theta}_n=\Theta\{\hat{C}_n|\mathbf{t}\}$, where $T_{\mathbf{d}}$ is the half-translation operator. Take $\bar{\Theta}$ and $\mathcal{G}_{\boldsymbol{\ell}}=\{\hat{m}_{\boldsymbol{\ell}}|\mathbf{t}\}$ for example, we study the evolution of the glide eigenvalues among the 8 time-reversal-invariant momenta (TRIM) $\boldsymbol{\Lambda}$ in the BZ. $\mathcal{G}_{\boldsymbol{\ell}}^2=-e^{-i2\mathbf{k}\cdot\mathbf{t}_{\parallel}}$, the glide eigenvalues for the mirror invariant plane is $g_{\pm}(\mathbf{k})=\pm ie^{-i\mathbf{k}\cdot\mathbf{t}_{\parallel}}$, and especially, $\mathcal{G}_{\boldsymbol{\ell}}^2(\boldsymbol{\Lambda})=\pm1$. $\bar{\Theta}^2=-e^{-i2\mathbf{k}\cdot\mathbf{d}}$ and $\bar{\Theta}^2(\boldsymbol{\Lambda})=\pm1$.
The relation $\bar{\Theta}\mathcal{G}_{\boldsymbol{\ell}}=e^{-i2\mathbf{k}_{\boldsymbol{\ell}}\cdot\mathbf{d}}
\mathcal{G}_{\boldsymbol{\ell}}\bar{\Theta}$, where $\mathbf{k}_{\boldsymbol{\ell}}$ is projection of $\mathbf{k}$ orthogonal to the mirror plane. $\gamma(\boldsymbol{\Lambda})\equiv e^{i2\boldsymbol{\Lambda}_{\boldsymbol{\ell}}\cdot\mathbf{d}}=\pm1$, therefore $\mathcal{G}_{\boldsymbol{\ell}}$ and $\bar{\Theta}$ always commute or anticommute with each other at TRIM.
$(\mathcal{G}_{\boldsymbol{\ell}}\bar{\Theta})^2=
e^{-i2\mathbf{k}_{\boldsymbol{\ell}}\cdot\mathbf{d}}\mathcal{G}_{\boldsymbol{\ell}}^2\bar{\Theta}^2$, thus $(\mathcal{G}_{\boldsymbol{\ell}}\bar{\Theta})^2=\pm1$ at $\boldsymbol{\Lambda}$. Now the 8 TRIM is classified into three cases, labelled as $\boldsymbol{\Lambda}_a,\boldsymbol{\Lambda}_b,\boldsymbol{\Lambda}_c$ if $\left(\bar{\Theta}^2,(\mathcal{G}_{\boldsymbol{\ell}}\bar{\Theta})^2\right)=(+1,+1),(\pm1,\mp1),(-1,-1)$. The Kramers degeneracy is guaranteed at $\boldsymbol{\Lambda}_b$ and $\boldsymbol{\Lambda}_c$, where the Kramers doublets are denoted as $(|\psi_{\mathbf{k}}\rangle,\bar{\Theta}|\psi_{\mathbf{k}}\rangle)$ if $\bar{\Theta}^2=-1$, or $(|\psi_{\mathbf{k}}\rangle,\mathcal{G}_{\boldsymbol{\ell}}\bar{\Theta}|\psi_{\mathbf{k}}\rangle)$ only when $(\mathcal{G}_{\boldsymbol{\ell}}\bar{\Theta})^2=-1$. Now the glide eigenvalues for the Kramers doublets are
$(g_1,g_2)_{\pm} = \pm(ie^{-i\boldsymbol{\Lambda}\cdot\mathbf{t}_{\parallel}},-i\gamma(\boldsymbol{\Lambda})e^{i\boldsymbol{\Lambda}\cdot\mathbf{t}_{\parallel}})$,
and $g_1/g_2=\gamma(\boldsymbol{\Lambda})\mathcal{G}^2_{\boldsymbol{\ell}}(\boldsymbol{\Lambda})=\pm1$. Therefore, $\gamma(\boldsymbol{\Lambda}_c)\mathcal{G}^2_{\boldsymbol{\ell}}(\boldsymbol{\Lambda}_c)=1$, $g_1(\boldsymbol{\Lambda}_c)=g_2(\boldsymbol{\Lambda}_c)\equiv g_c$; and $\gamma(\boldsymbol{\Lambda}_b)\mathcal{G}^2_{\boldsymbol{\ell}}(\boldsymbol{\Lambda}_b)=-1$, $g_1(\boldsymbol{\Lambda}_b)=-g_2(\boldsymbol{\Lambda}_b)\equiv g_b$. Then we consider a path $\mathcal{C}$ connecting $\boldsymbol{\Lambda}_c$ to $\boldsymbol{\Lambda}_b$ on the mirror invariant plane in the BZ, the glide eigenvalues of the Kramers doublets should smoothly evolve from $\pm(g_c,g_c)$ to $(g_b,-g_b)$ as shown in Fig.~\ref{fig1}(b). In the absence of other degeneracies, there must be a band crossing at certain point $\mathbf{k}_{\mathcal{C}}$ on $\mathcal{C}$, where the Kramers pairs switch partners. Since the band crossing is true for arbitrary $\mathcal{C}$ between $\boldsymbol{\Lambda}_c$ to $\boldsymbol{\Lambda}_b$, a NL is formed at $\mathbf{k}_{\mathcal{C}}$. Specifically, such NL is essential, which cannot be gapped without lowering the magnetic SG symmetries. Also, it is filling-enforced, for the bands shown in Fig.~\ref{fig1}(b), the NL must exist at the Fermi level when filling $\nu\in4\mathbb{Z}+2$. Furthermore, if the system has $\mathcal{I}$ or $\mathcal{I}\Theta$ symmetry, each band must be doubly degenerate, then the crossing point $\mathbf{k}_{\mathcal{C}}$ will shift to TRIM, leading to an essential DP.

\begin{figure}[t]
\begin{center}
\includegraphics[width=3.3in]{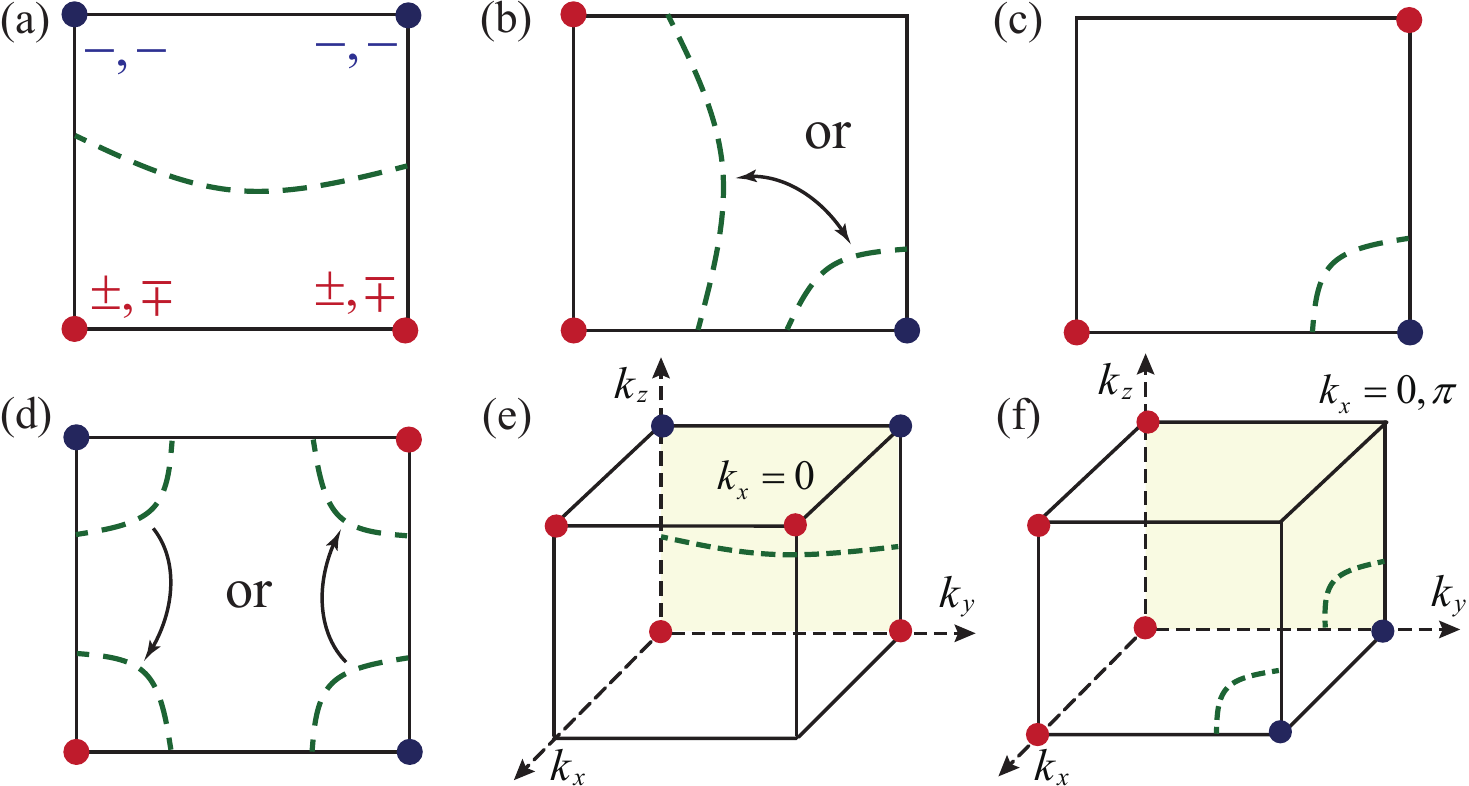}
\end{center}
\caption{(color online). (a)-(d) Four representative examples of NLs (indicated by green dashed lines) on the mirror-invariant plane. The square is just $1/4$ of 2D BZ, and the corners denotes TRIM. Red and blue dots have two-fold degeneracy with the opposite and same glide eigenvalues, respectively. There are two possible NL topology in (b) \& (d). Two representative cases of line nodes in the BZ. (e) Line node is at $k_x=0$ only. Here $\mathbf{d}=(\frac{1}{2}00)$, $\mathbf{t}=(00\frac{1}{2})$. (f) At both $k_x=0,\pi$. Here $\mathbf{d}=(00\frac{1}{2})$, $\mathbf{t}=(0\frac{1}{2}0)$. The complete lists of NLs in the BZ are shown in the Supplemental Material~\cite{supplementary}.}
\label{fig3}
\end{figure}

The remaining question is whether such path $\mathcal{C}$ exists. We consider $\boldsymbol{\ell}=\hat{x}$ for concreteness. The positions and topology of the NLs can be determined by studying the values of $(\bar{\Theta}^2,(\mathcal{G}_x\bar{\Theta})^2)$ at TRIM in the mirror invariant plane $k_x=0,\pi$.
Four representative examples are shown in Fig.~\ref{fig3}(a)-(d). There are only two possible cases of NLs in the BZ listed in Fig.~\ref{fig3}(e) and~\ref{fig3}(f), which reside at $k_x=0$ only and at both $k_x=0,\pi$, respectively, depending on the value of $\boldsymbol{\ell}\cdot\mathbf{t}$ when $\mathbf{d}_{\parallel}=0$. It is worth mentioning that the glide mirror plane for essential NLSM must have $\mathbf{t}_{\parallel}\neq0$~\cite{supplementary}, which is quite different from band-inversion NLSM with $\boldsymbol{\ell}\cdot\mathbf{t}\neq0$.

More generally, an essential NL always appears on the mirror invariant plane connecting either two doubly degenerate high-symmetry points or high-symmetry lines, or high-symmetry points and high-symmetry lines, where the glide eigenvalues for the degenerate two bands are the same at one point or line and opposite at the other point or line. The two-fold degeneracy could originate from an antiunitary operator such as $\bar{\Theta}$ discussed above or noncommutativity of two unitary operators~\cite{murakami2017}. Similarly, the antiunitary operator $\bar{\Theta}_n=\Theta\{\hat{C}_n|\mathbf{t}\}$ could also give rise to the two-fold degeneracy, which can be obtained by analyzing the irreducibility of a corepresentation of a magnetic SG through Herring rule~\cite{bradley1972}. 2D irreducible corepresentation exsits for $n=4,6$~\cite{zhang2015}. The above analysis also applies to NLs in 2D.

\begin{figure}[t]
\begin{center}
\includegraphics[width=3.3in]{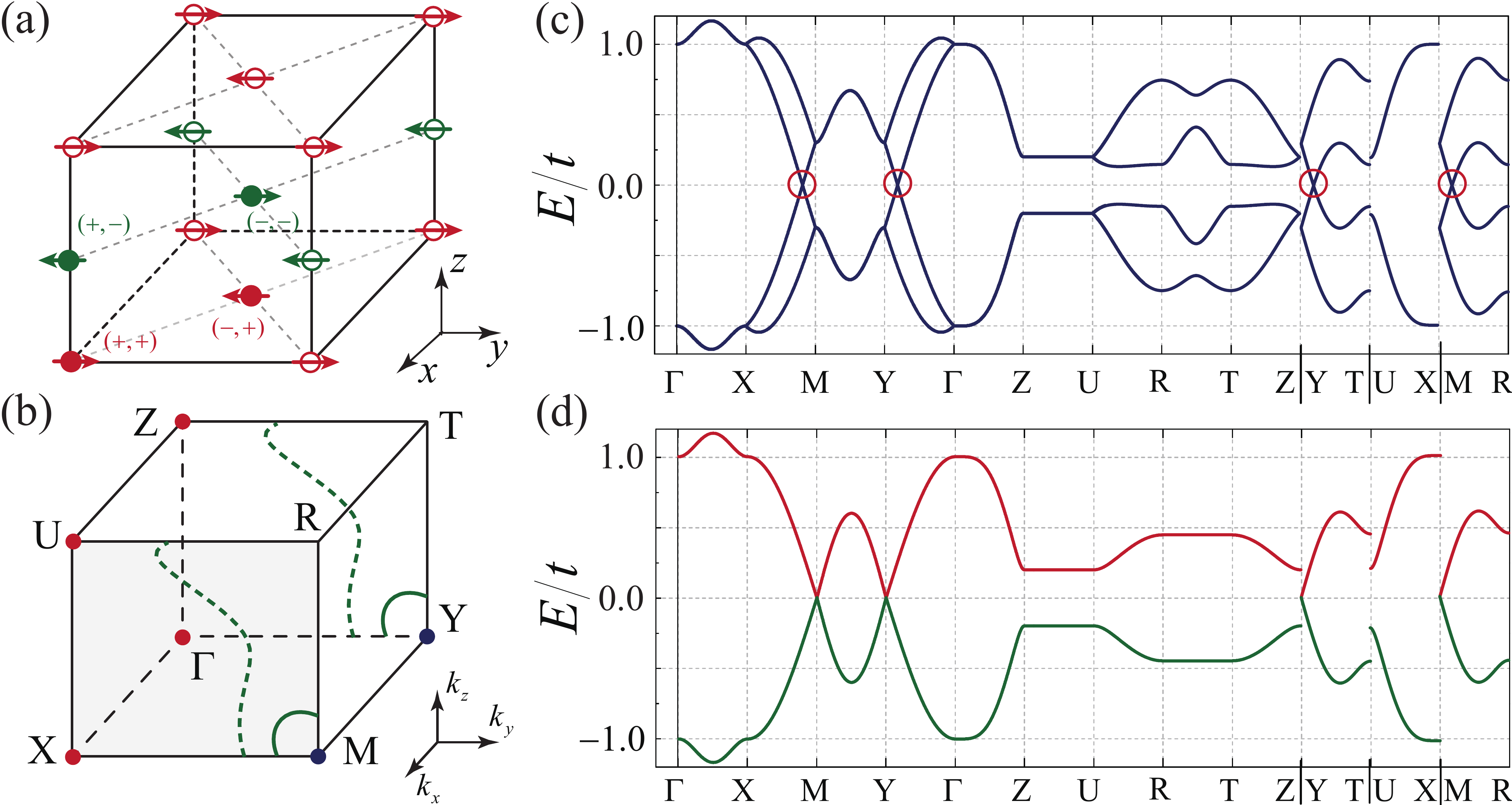}
\end{center}
\caption{(color online). (a) The orthorhombic primitive lattice structure of SG~28.
(b) Energy band described by the model of Eq.~(\ref{model2}), with $t=1.0$, $t_1=-0.4$, $\lambda_1=0.2$, $\lambda_2=0.6$, $\lambda_3=0.5$, and $\lambda_4=0.3$. Only the upper subsystem is shown. The 2D DPs are indicated by red circles. (c) The essential line nodes (indicated by green lines) in the BZ, where the position changes for $\lambda_4=0.3$ (solid) and $\lambda_4=0.6$ (dashed). (d) The essential AFM DSM in 3D with $\lambda_4=0$, where the two DPs need not be at the same energy.}
\label{fig4}
\end{figure}

We construct a tight-binding model which exhibits essential NL. In Fig.~\ref{fig4}(a), the lattice has an orthorhombic primitive lattice structure of SG~28 (layer group $Pma2$). This can be viewed as stacking the 2D crinkled lattice in Fig.~\ref{fig2}(b) along $z$-axis, but with opposite magnetic order in the two adjacent layers. With further $\mathcal{I}$ breaking and the $\pm\hat{y}$ direction AFM ordering, the symmetry of SG~59 in the paramagnetic state will reduce to SG~28. The unit cell contains four sublattices, indexed by $(\tau_z,\sigma_z)$ associated with the basis vectors $\mathbf{t}_0=\frac{1}{2}[(1-\tau_z)(c'\frac{1}{2}c)+(1-\sigma_z)(00\frac{1}{2})]$, where $0<c',c<\frac{1}{2}$. Each lattice site contains an $s$ orbital, which in general leads to an 8-band model. The system respects $\bar{\Theta}$ with $\mathbf{d}=(00\frac{1}{2})$. The symmetry generators and their representations in the sublattice space are $\bar{\Theta}=i\tau_y\mathcal{K}\sigma_z$ and $\{\hat{m}_x|0\frac{1}{2}0\}=\tau_z\sigma_y$. We then assume AFM interaction is much stronger than hopping and SOC terms, therefore the system is effectively decoupled into two 4-band models as time-reversal partners, each with one spin per sublattice. The upper subsystem consists of two pairs of sublattices index by $\tau_z$, where each pair respects $\bar{\Theta}$ and is related to each other by $\{\hat{m}_x|0\frac{1}{2}0\}$.
The projection of the 8-band model to the upper subsystem leads to the simplified Hamiltonian
\begin{eqnarray}\label{model2}
\mathcal{H}_b &=& t\tau_x\left(\sigma_x\cos\frac{k_x}{2}+\sigma_y\sin\frac{k_x}{2}\right)\cos\frac{k_y}{2}\cos\frac{k_z}{2}
\nonumber
\\
&&+t_1\tau_x\left(\sigma_x\cos\frac{k_x}{2}+\sigma_y\sin\frac{k_x}{2}\right)\sin\frac{k_y}{2}\sin\frac{k_z}{2}
\nonumber
\\
&&+\lambda_1\tau_x\sigma_z\sin\frac{k_z}{2}+\lambda_2\tau_y\cos\frac{k_z}{2}\sin k_x+\lambda_3\tau_z\sin k_z
\nonumber
\\
&&+\lambda_4\tau_z\left(\sigma_x\sin\frac{k_x}{2}+\sigma_y\cos\frac{k_x}{2}\right)\sin\frac{k_y}{2}.
\end{eqnarray}
Here $t_i$ describes the hopping, $\lambda_i$ is SOC. $\lambda_4$ is the only term which breaks $\mathcal{I}\Theta$. If $\lambda_4=0$, with $\bar{\Theta}$, $\{\hat{m}_x|0\frac{1}{2}0\}$ and $\mathcal{I}\Theta$ symmetries, the system is an essential AFM DSM in 3D~\cite{wang2017b} with two symmetry-inequivalent DPs located at $M$ and $Y$ as shown in Fig.~\ref{fig4}(d). In Fig.~\ref{fig4}(b) and~\ref{fig4}(c), with $\lambda_4\neq0$, the 3D DP evolves into NL enclosing a TRIM, consistent with the analysis in Fig.~\ref{fig3}(f).

\emph{Discussion.}
We briefly discuss the surface state in AFM NLSMs. For band-inversion NLSM, the protection of NLs requires $\mathcal{I}\Theta$. However, the surface of a physical system always breaks $\mathcal{I}\Theta$ and leads to gapped or no protected surface states. While for essential NLSM, the protection of NLs is guaranteed by $\bar{\Theta}$ and $\{\hat{m}_{\boldsymbol{\ell}}|\mathbf{t}\}$. The open surface with both symmetries will lead to non-trivial topological surface states, for example on the (100) surface of the model in Eq.~(\ref{model2}).
The surface band and AFM fluctuations may provide a unique platform for interesting strong correlated physics.

Furthermore, we briefly discuss the search principle for realistic materials and comment on the possible candidate. From the model presented above, aside from appropriate crystalline symmetry, one can see that band-inversion NLSM may exist in materials with strong SOC and relatively weak AFM interaction. Especially, finding 3D band-inversion NLSM is reduced to searching for 2D AFM DSM in layered materials~\cite{wang2017a}. The tetragonal AFM CuMnAs breaks both $\mathcal{I}$ and $\Theta$ whereas $\mathcal{I}\Theta$ holds. The Mn lattice is similar to the structure in Fig.~\ref{fig2}(a), which determines the low energy electronic structure. Interestingly, the N\'eel vector direction is electrically controllable~\cite{wadley2016}, when $\hat{\mathbf{n}}\parallel[100]$, it should realize band-inversion NLs. Unlike band-inversion NLSM, essential magnetic NLSM only requires specific magnetic SG symmetry, which is compatible with the narrow band width in $d$-orbital. From the model in Eq.~(\ref{model2}), one can see that essential NLSM can emerge as an $\mathcal{I}$ breaking phase from 3D essential magnetic DSM with glide mirror symmetry.

In summary, we extend the topological NLSMs in $\Theta$-invariant systems to that in magnetic systems, which is based on the irreducible corepresentation of magnetic SGs. Two distinct classes of NLSMs have been identified, where the corresponding accidental and essential NLs may have different magneto-transport properties. Furthermore, the magnetic fluctuations exists generically, which may lead to a dynamical axion field~\cite{li2010,wang2016a}. The magnetic NLSMs proposed here may provide a platform for the interplay between magnetism and exotic topological phases.

\begin{acknowledgments}
This work is supported by the Natural Science Foundation of Shanghai under Grant No.~17ZR1442500; the National Thousand-Young-Talents Program; the National Key Research Program of China under Grant No.~2016YFA0300703; the Open Research Fund Program of the State Key Laboratory of Low-Dimensional Quantum Physics, through Contract No.~KF201606; and by Fudan University Initiative Scientific Research Program.
\end{acknowledgments}

\end{document}